\documentclass[12pt]{iopart}

\usepackage{iopams}  
\usepackage[dvipdfmx]{graphicx}
\expandafter\let\csname equation*\endcsname\relax
\expandafter\let\csname endequation*\endcsname\relax
\usepackage{amsmath}
\usepackage{physics}
\usepackage{textcomp}
\usepackage{color}

\begin{document}

\title[]{Resistivity measurement for non-magnetic materials using
high-order resonance mode of MFM-cantilever oscillation}

\author{Kazuma Okamoto$^1$, Takumi Imura$^1$, Satoshi Abo$^1$, Fujio Wakaya$^1$, Katsuhisa Murakami$^2$, Masayoshi Nagao$^2$}

\address{$^1$ Graduate School of Engineering Science, Osaka University, 1-3 Machikaneyama, Toyonaka, Osaka 560-8531, Japan}
\address{$^2$ National Institute of Advanced Industrial Science and Technology, 1-1-1 Umezono, Tukuba, Ibaraki 305-8560, Japan}
\ead{u329610i@ecs.osaka-u.ac.jp}
\vspace{10pt}
\begin{indented}
\item[]\today
\end{indented}

\begin{abstract}
A method to measure the electrical resistivity of materials using magnetic-force microscopy (MFM) is 
discussed, where MFM detects the magnetic field caused by the tip-oscillation-induced eddy current.  
To achieve high sensitivity, a high cantilever oscillation frequency is preferable, 
because it induces large eddy currents in the material. 
Higher-order resonance modes of the cantilever oscillation leads to higher frequency. 
To discuss such high-order-mode oscillation, 
a differential equation governing MFM cantilever oscillation in the high-order resonance mode 
is formulated, and 
an analytical solution of the phase difference is obtained.  
The result shows that the phase difference decreases at higher modes, because   
the effective spring constant increases faster than the force from the 
eddy current.  
\end{abstract}

%
%
%
%
%

\section{Introduction}
The impurity density or resistivity of semiconductors is a significant parameter for semiconductor devices\cite{dk}. 
Several techniques are used to dope semiconductors with impurities, such as 
thermal diffusion, ion implantation, plasma doping, epitaxial growth, $etc$\cite{cz,cx,da}. 
To activate the impurities, annealing is generally used, during which the impurity atoms may be displaced.  
Measurement techniques are essential for improving the quality and stable production of semiconductor devices\cite{bp,db,dc}. 
Scanning spreading resistance microscopy (SSRM) is such a technique, which evaluates the 
spreading resistance via current measurements with an applied voltage, and can provide the 
impurity density with high spacial resolution\cite{cx}. However, the reproducibility is poor due to 
scratching of the sample surface\cite{cv}.  
In this work, an impurity density measurement method is discussed using magnetic-force microscopy (MFM). MFM is generally used to measure stray magnetic fields of microscopic magnetic domain structures, magnetic recording mediums, \textit{etc.}, where a magnetized tip oscillating near the sample surface is used\cite{an,dd,br}. 
The reproducibility of MFM measurements is better than that of SSRM, because the tip and sample surface are 
not touched nor biased\cite{df}.  

It has been reported that MFM can detect signals from non-magnetic materials\cite{ci}. 
Specifically, MFM detects the magnetic field generated by the eddy current caused by the oscillation of the magnetized MFM tip\cite{n}. 
The signal proportional to the derivative of magnetic field is detected as the phase difference in the MFM-tip oscillation. The resistivity of the material can thus be calculated from the phase difference\cite{n}. 
Because the resistivity of semiconductors is significantly higher than that of metals\cite{cs,af},  
the sensitivity of MFM measurements should be improved for measuring semiconductors\cite{n}. 
In high-oder resonance modes at high frequencies, 
the eddy current should increase.  
High-order resonance modes should therefore be advantageous in realizing 
high-sensitivity measurements using MFM.
The purpose of this work is to provide a theoretical expression of the phase difference 
in resistivity measurements using MFM with high-order resonance modes.

\section{\label{sec:level1}Free oscillation of the cantilever}
\begin{figure}
\centering
\includegraphics[width=80mm]{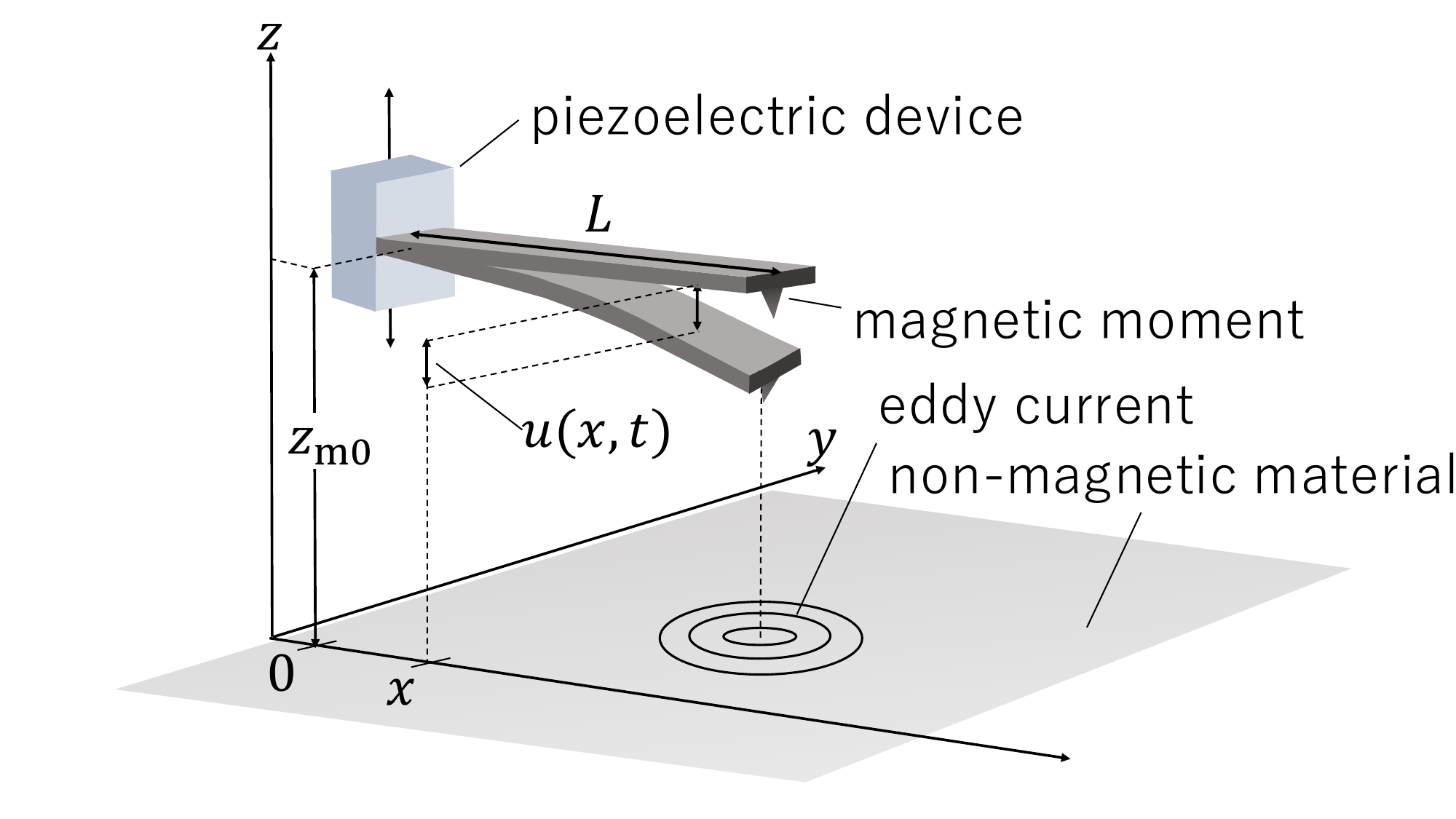}
\caption{Schematic of MFM system measuring the resistivity measurement of non-magnetic materials.
$L$ is the length of the cantilever, 
$u(x,t)$ is the displacement of the cantilever at position $x$ and time $t$, and 
$z_{\rm m0}$ is the average height of the cantilever from the material surface. 
}
\label{fig:model}
\end{figure}

Figure \ref{fig:model} shows a schematic of an MFM system measuring the resistivity 
of non-magnetic materials, 
where 
$L$, $u(x,t)$, and $z_{\rm m0}$ denote the length of the cantilever,  
displacement of the cantilever at position $x$ and time $t$,    
and average height of the cantilever from the material surface, respectively. 

The forced oscillation should be formulated using 
the analytical solutions for the free oscillation 
without excitation, damping and external forces.  
The analytical solutions for the free oscillation and their orthogonality are discussed in this section.

\subsection{\label{sec:level2} {\rm Analytical solution for free oscillation}}
\begin{figure}[t]
\centering
\includegraphics[width=85mm]{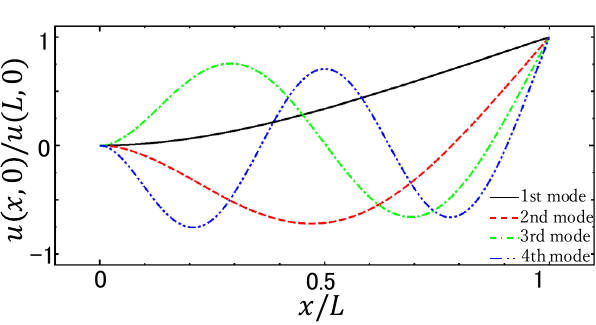}
\caption{\label{fig:modesinpuku}Cantilever displacement as expressed by Eq. (2) at $t=0$ in each eigenmode normalized 
by the displacement at $x=L$. }
\end{figure}

The equation of the cantilever in free oscillation is\cite{x}
\begin{align}
EI\frac{\partial^4u(x,t)}{\partial x^4}+\rho_{\rm d} S\frac{\partial^2u(x,t)}{\partial t^2}=0,\label{eq:banehouteisiki}
\end{align}
where $E$ is the Young's modulus, 
$I$ is the cross-sectional inertia, 
$\rho_{\rm d}$ is the density of the cantilever material and 
$S$ is the cross-sectional area of the cantilever in the $y$-$z$ plane.  
One of the solutions of  Eq. (\ref{eq:banehouteisiki}) 
with free-oscillation boundary conditions\cite{az}, 
$u(0,t)=0$, $\pdv{u(0.t)}{x}=0$, $\pdv[2]{u(L,t)}{x}=0$, and $\pdv[3]{u(L,t)}{x}=0$, 
is\cite{x,az,ai}
\begin{align}
&u(x,t)=C{\Psi}(x)T(t) \label{eq:777}, \\
&\Psi(x)=\qty\big(\sin \lambda x-\sinh \lambda x) + \Xi\qty\big(\cos  \lambda x -\cosh  \lambda x),
\\
&T(t) = A_{0}\cos\omega t,\\
&\Xi \equiv \frac{\cos\lambda L+\cosh\lambda L}{\sin\lambda L-\sinh\lambda L} ,\\
& \lambda=\sqrt[4]{\frac{\rho_\mathrm{d} S}{EI}\omega^2}, \label{eq:6}
\end{align}
where $A_{0}$ and $C$ are constants.  
The eigenequation for $\lambda$  
\begin{align}
1+\cos\lambda L\cosh\lambda L=0  \label{eq:lambda} 
\end{align}
leads to the eigenvalues labeled as $\lambda_i$ ($i=1,2,3,\cdots$) in order from the smallest as 
\begin{equation}
\lambda_i L = 1.875, 4.694, 7.854, 10.995 \cdots, 
\label{eq:8}
\end{equation}
and corresponding $\omega_i$ using Eq. (\ref{eq:6}).  
%
Figure \ref{fig:modesinpuku} shows the eigenmodes of cantilever oscillation 
as calculated above. 
It can be confirmed that the functions $\Psi_{i}(x)$ satisfies  
\begin{equation}
EI\dv[4]{\Psi_{i}(x)}{x}-\rho_{\rm d}S\omega^2_{i}\Psi_{i}(x)=0, \label{eq:4to0}
\end{equation}
which will be used in the following section for changing 
$\dv[4]{\Psi_i(x)}{x}$
to $\Psi_i(x)$.

\subsection{\label{4}{\rm Mode orthogonality}}
The functions $\Psi_{i}(x)$ satisfy the orthogonality relations\cite{az} below and 
 make a complete system concerning $x$, 
\begin{align}
\rho_{\rm d}S\int^{L}_{0}\Psi_{n}(x)\Psi_{i}(x)dx
=\left\{
\begin{array}{l}
0\;\;\;\;\;(i\neq n)\\
M_{i}\;\;(i=n),\label{eq:tyokkou}
\end{array}
\right.
\end{align}
where $M_{i}$ is called the mode mass.
The values of $M_i$ are 
$ 9.53\times 10^{-11}, 4.95 \times 10^{-11}, 5.15 \times 10^{-11}, \cdots$, kg; 
when 
$S=9.8\times10^{-11}$, $\rho_\mathrm{d} = 2.33$ g/cm$^3$,
as discussed in \S 4.

\section{\label{cc}Damped and forced oscillation of the cantilever}
The equation of motion of the cantilever with damping and external forces is modified 
from Eq. (\ref{eq:banehouteisiki}) as\cite{bc}
\begin{align}
EI\frac{\partial^4u(x,t)}{\partial x^4}+\eta I\frac{\partial^5u(x,t)}{\partial t\partial x^4}+\rho_{\rm d} S\frac{\partial^2u(x,t)}{\partial t^2}=F(x,t),\label{eq:katamotimfm}
\end{align}
where $\eta$ is the viscosity coefficient and $F(x,t)$ is the distributed external force
per unit length. 
\subsection{{\rm External force}}
The external force $F(x,t)$ in Eq. (\ref{eq:katamotimfm}) comprises 
the excitation force  $F_{1}(x,t)$, which is applied to the cantilever by oscillating the piezoelectric device 
and the force from the magnetic field  $F_{2}(x,t)$. 
The excitation force $F_{1}(x,t)$ can be written as\cite{bc}
\begin{align}
F_{1}(x,t)&=F_{\mathrm{e}}\delta(x-L)\cos\omega_\mathrm{F} t,\label{eq:kasin}
\end{align}
where $F_\mathrm{e}$ ($ > 0$ ), \textcolor{red}{$\delta(x)$} and $\omega_\mathrm{F}$ are the constant force, \textcolor{red}{delta function} and angular frequency of the forced oscillation, respectively.  
The force from magnetic field $F_{2}(x,t)$ can be written as\cite{n}
\begin{align}
F_{2}(x,t)&=-\frac{3}{64\pi}\frac{p^2}{\rho}\frac{\delta(x-L)}{\{z_{\rm m0}+u(L,t)\}^4}\frac{\partial u(L,t)}{\partial t},\label{eq:zibaziba}
\end{align}
where $\rho$ and $p$ are the sheet resistivity of the non-magnetic material and magnetic moment of the 
MFM tip, respectively. 
Using Eq. (\ref{eq:zibaziba}), Eq. (\ref{eq:katamotimfm}) becomes  a non-linear differential equation, 
meaning that it is difficult to find analytical solutions. 
If $z_{\rm m0}\gg u(L,t)$, 
Eq. (\ref{eq:zibaziba}) becomes 
\begin{align}
F_{2}(x,t)&\cong-\frac{3}{64\pi}\frac{p^2}{\rho}\frac{\delta(x-L)}{z_{\rm m0}^4}\frac{\partial u(L,t)}{\partial t}.
\label{eq:12}
\end{align}
Equations (\ref{eq:katamotimfm}), (\ref{eq:kasin}) and (\ref{eq:12}) give the final equation for 
the system with the excitation force, damped force, and the force from the eddy current as 
\begin{align}
&EI\frac{\partial^4u(x,t)}{\partial x^4}+\eta I\frac{\partial^5u(x,t)}{\partial t\partial x^4}+\rho_{\rm d} S\frac{\partial^2u(x,t)}{\partial t^2}\nonumber\\
&=-\frac{3}{64\pi}\frac{p^2}{\rho}\frac{\delta(x-L)}{z_{\rm m0}^4}\frac{\partial u(L,t)}{\partial t}+F_{a}\delta(x-L)\cos\omega_\mathrm{F} t. 
\label{eq:tuini}
\end{align}
The stationary solution of this equation gives  
the oscillation phase of the cantilever tip end, 
which is observed in the actual experiment.

\subsection{{\rm Analytical solution for damped and forced oscillations}}
Using the complete functions $\Psi_{i}(x)$,
the solution of Eq. (\ref{eq:tuini}) with the boundary condition for damped and forced oscillation 
can be expressed as\cite{bg,be}
\begin{equation}
u(x,t)=\sum_{i=1}^{\infty}c_{i}(t)\Psi_{i}(x).\label{eq:sindoukai2}
\end{equation}
To find the phase difference in the stationary state, 
it is required to solve the equation concerning the expansion coefficient $c_{i}(t)$.  
Using Eqs.  (\ref{eq:4to0}), (\ref{eq:tyokkou}), (\ref{eq:tuini}), and (\ref{eq:sindoukai2}), 
the differential equation for $c_{i}(t)$ is derived as 
\begin{align}
\dv[2]{c_n(t)}{t}&+\nu_{n}\dv{c_{n}(t)}{t}+\omega^2_{n}c_{n}(t)=\frac{F_{a}{\Psi}_{n}(L)}{M_{n}}\cos\omega_\mathrm{F} t,\label{eq:mode}
\end{align}
where the effective viscosity coefficient $\nu_{n}$ and characteristic tip height $z_{\rm c}$ are respectively defined by
\begin{align}
\nu_{n}&\equiv\frac{\eta}{E}\omega^2_{n}+\frac{z^4_{\rm c}}{z^4_{m0}}\omega_{n}\\
z^4_{\rm c}&\equiv\frac{3}{64\pi}\frac{p^2{\Psi}_{n}^2(L)}{\rho M_{n}\omega_{n}}.\label{eq:gamma'}
\end{align}
The stationary solution of Eq. (\ref{eq:mode}) is 
\begin{align}
& c_{n}(t)=A_{n}\cos(\omega_\mathrm{F} t+\phi_{n}), \label{eq:cc}  \\
& A_{n}=\frac{F_\mathrm{e}\abs{\Psi_{n}(L)}}{M_{n}\sqrt{(\omega^2_{n}-\omega_\mathrm{F}^2)^2+(\nu_{n}\omega_\mathrm{F})^2}}, \label{eq:19}\\
&\phi_{n}=-\atan\left[\frac{\nu_{n}\frac{\omega_\mathrm{F}}{\omega^2_{n}}}{1-(\frac{\omega_\mathrm{F}}{\omega_{n}})^2}\right]-\pi\theta\left(\frac{\omega_\mathrm{F}}{\omega_{n}}-1\right),  
\label{eq:nnn}
\end{align}
where $\theta(x)$ is the step function. 
As above, the solution of Eq. (\ref{eq:tuini}) is expressed as the summation of all eigenmodes.  
Around the resonance frequency of the $n$th mode $\omega_{n}$, 
the amplitudes of the other modes are  substantially smaller than that of the $n$th mode. 
Therefore, if $\omega_\mathrm{F} \simeq \omega_n$, 
the displacement of the cantilever is approximated as $u(x,t) \simeq u_{n}(x,t)$.
In the typical MFM experiment, the phase difference of the cantilever oscillation  at $x=L$ is detected 
using laser light, which is the same as $\phi_n$.

\section{Actual values of parameters in the equations} 

To discuss the theoretical results in the previous sections, 
it is helpful to express them graphically, where 
the actual values of the parameters in the equations should be fixed.  

The MFM cantilever used in previous experiments\cite{n} 
was commercially available as model\# MESP provided by Bruker Co., with dimensions 
$225{\rm \mu m}\times35{\rm \mu m}\times2.8{\rm \mu m}$, resulting in $I= 64.0$ $\mu$m$^4$.  

The cantilever is made from silicon, with material parameters 
$\rho_\mathrm{d} = 2.33$ g/cm$^3$, $E=179$ GPa\cite{di,dh}.  
The angular frequency of each mode $\omega_n$ is determined by the dimensions and 
material parameters with Eqs.(6) and (8) 
as
$5.03 \times 10^5$, $3.15\times 10^6$, 
$8.82 \times 10^6$, $1.73 \times 10^7$, $\cdots$, rad/s.

The magnetic moment of the MFM tip is typically $p=1.0\times 10^{-18}$ Wb$\cdot$m\cite{bo}.  
The sheet resistivity of the non-magnetic metal with thichkness of $\sim 1$ nm is 
$\rho=10$ $\Omega/\mathrm{sq.}$\cite{dj}

\section{Relationship between the Q-value and viscosity coefficient}
\begin{figure}
\centering
\includegraphics[width=75mm]{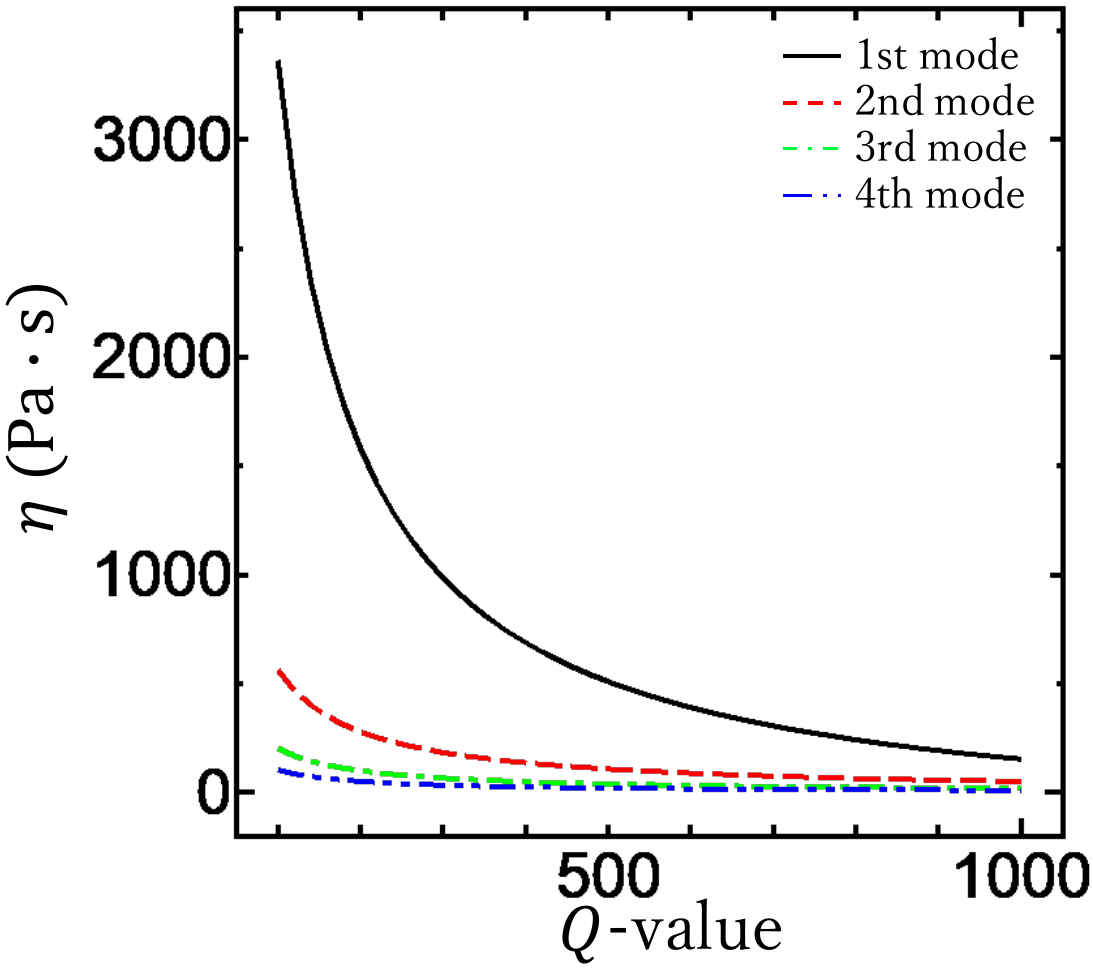}
\caption{\label{fig:Qgamma}The viscosity coefficient as a function of the Q-value as 
expressed by Eq. (26). $z_{\rm m0}=30\;{\rm nm}$, $E=179\;{\rm GPa}$, $L=225\;\mu {\rm m}$, $p=10^{-18}\;{\rm Wb\cdot m}$, $S=9.8\times10^{-11}\;{\rm m^2}$,  $I= 64.0$ $\mu$m$^4$, $\rho=10\;\Omega/{\rm sq.}$, $\rho_{\rm d}=2.33\;{\rm g/cm^3}$.}
\end{figure}

The Q-value is related to the viscosity coefficient $\eta$; 
increasing the Q-value decreases $\eta$ 
and improves the sensitivity\cite{av}. The Q-value is defined by\cite{cu}
\begin{equation}
Q\equiv2\pi\frac{W_{\rm acc}}{W_{\rm out}},\label{eq:Qteigi}
\end{equation}
where $W_{\rm acc}$ is the accumulated energy in the system and 
$W_{\rm out}$ is the energy lost per period.
In the dumped and forced oscillation systems, 
the energies $W_{\rm acc}$ and $W_\mathrm{out}$ are\cite{az,bc} 
\begin{align}
W_{\rm acc}&=\int^{L}_{0}\frac{EI}{2}\left(\frac{\partial^2u(x,t)}{\partial x^2}\right)^2{\rm d}x+\int^{L}_{0}\frac{1}{2}\rho_{\rm d}S\left(\frac{\partial u(x,t)}{\partial t}\right)^2{\rm d}x,  \nonumber\\
W_{\rm out}&=\int^{\frac{2\pi}{\omega}}_{0}\int^{L}_{0}F_\mathrm{e}\delta(x-L)\cos\omega_\mathrm{F} t\frac{\partial u(x,t)}{\partial t}{\rm d}x{\rm d}t.  \nonumber
\end{align}
If $\omega_\mathrm{F} \simeq \omega_n$, 
\begin{align}
&W_\mathrm{acc}
\simeq\frac{1}{2}C^2M_{n}\omega^2_{n}A_{n}^2, 
\label{eq:naibu} \\
&W_\mathrm{out}
\simeq\frac{\pi F_\mathrm{e}A_{n}\nu_{n}\omega_\mathrm{F} C{\Psi}_{n}(L)}{\sqrt{(\omega^2_{n}-\omega_\mathrm{F}^2)^2+(\nu_{n}\omega_\mathrm{F})^2}}. 
\label{eq:WWW}
\end{align}
Using Eqs. (\ref{eq:19}), (\ref{eq:Qteigi}), (\ref{eq:naibu}), and (\ref{eq:WWW}), Q-value of each mode is derived as 
\begin{align}
Q_{n}=\frac{\omega_{n}^2}{\left(\frac{\eta}{E}\omega^2_{n}+\frac{z^4_{\rm c}}{z^4_{\rm m0}}\omega_{n}\right)\omega_\mathrm{F}}.\label{eq:Q}
\end{align}
The viscosity coefficient $\eta$ cannot be determined directly from experiments.  
The parameter related to $\eta$ that can be determined in the actual experiment is the 
Q-vlaue.  
The Q-value obtained from experiments in the air is typically 100--1000, depending on $n$\cite{dg}.    
Using Eqs. (\ref{eq:6}) and (\ref{eq:8}) with material and dimension parameters $\rho_\mathrm{d}$, 
$S$, $E$ and  $I$, the mode angular frequency $\omega_n$ is fixed.  
Moreover, using the sheet resistivity $\rho$ and the magnetic moment $p$,  
the characteristic tip height $z_\mathrm{c}$ is also fixed.   
If $z_\mathrm{m0}$ and $\omega_\mathrm{F}$ are fixed from the typical experimental setup,  
Eq. (\ref{eq:Q}) gives $\eta$ as a function of $Q_n$, as shown in Fig. \ref{fig:Qgamma}, 
which should be denoted as $\eta_n$\cite{bm}.    

\section{Phase difference at the cantilever end}
\subsection{{\rm Phase difference using independent parameters}}
Equation (\ref{eq:nnn}) gives the phase at the average cantilever height $z_{\rm m0}$. 
When  $z_{\rm m0}\rightarrow\infty$, the phase is uninfluenced by the non-magnetic material and given 
as  
\begin{align}
\phi_{n}(\infty)=-\atan\left[\frac{\eta_{n}\omega^2_{n}}{E}\frac{\frac{\omega_\mathrm{F}}{\omega^2_{n}}}{1-(\frac{\omega_\mathrm{F}}{\omega_{n}})^2}\right]-\pi\theta\left(\frac{\omega_\mathrm{F}}{\omega_{n}}-1\right).\label{eq:infty}
\end{align}
Using Eqs. (\ref{eq:nnn}) and (\ref{eq:infty}), the phase difference is expressed as 
\begin{align}
&\phi_{n}(z_{\rm m0})-\phi_{n}(\infty)\nonumber\\
&=-\atan\left[\frac{Ez^4_{\rm c}\Delta_{n}\left(1-\Delta_{n}^2\right)}{z_{\rm m0}^4\left\{E\left(1-\Delta_{n}^2\right)^2+\eta_{n}\Delta_{n}^2\left(\frac{\eta_{n}}{E}\omega^2_{n}+\frac{z^4_{\rm c}}{z^4_{\rm m0}}\omega_{n}\right)\right\}}\right],\label{eq:uuu}
\end{align}
where $\Delta_{n}\equiv\omega_{\rm F}/\omega_{n}$. 
Using Eq. (\ref{eq:Q}), Eq. (\ref{eq:uuu}) is rewritten as 
\begin{align}
\phi_{n}(z_{\rm m0})-\phi_{n}(\infty)=-\atan\left[\frac{Q_n^2z^4_{\rm c}\Delta_{n}\left(1-\Delta_{n}^2\right)}{z_{\rm m0}^4\left\{Q_n^2\left(1-\Delta_{n}^2\right)^2+1\right\}-Q_nz^4_{\rm c}\Delta_{n}}\right].\label{eq:isousadokurituhizisei}
\end{align}
Note that Eq. (29) is more useful than Eq. (28) because $Q_n$ can be determined by experiments. 
\subsection{{\rm Discussion}}
\begin{figure}[t]
\centering
\includegraphics[width=70mm]{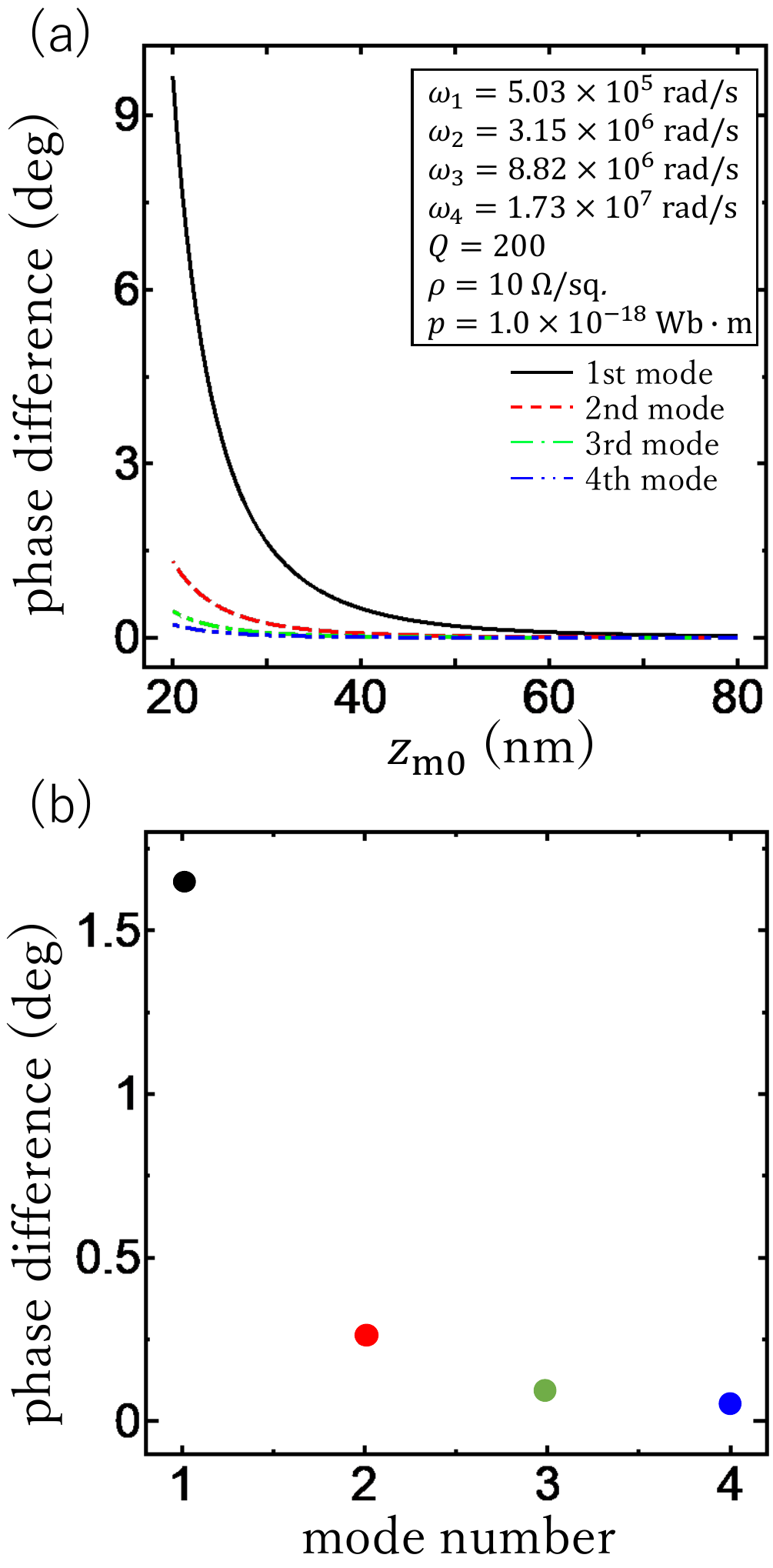}
\caption{\label{fig:takasa}
Phase difference with   $\Delta_n=0.998$ expressed as  Eq.\ (\ref{eq:isousadokurituhizisei}) as a function of 
(a) $z_{\rm m0}$ and  
(b)mode number with $z_\mathrm{m0} = 30$ nm.}
\end{figure}

The phase differences expressed as Eq.\ (\ref{eq:isousadokurituhizisei}) are plotted in Fig.\ \ref{fig:takasa}(a) as a function of $z_\mathrm{m0}$, 
where $Q_n$ is fixed as 200\cite{bo,cw}. 
The Q-values of the $n$th mode are generally different for different $n$ and can be determined 
from the experimentally observed frequency dependence of the oscillation amplitude\cite{df}.   
However, all Q-values are fixed as 200 in Fig. \ref{fig:takasa}, because 
the corresponding theoretical expressions are not provided.  
The phase difference depends on $\Delta_{n}$, 
which is fixed in Fig.~\ref{fig:takasa} so that the 
phase difference becomes maximum in each mode.  
The phase difference shown in Fig. \ref{fig:takasa}(a) decreases with increasing $z_\mathrm{m0}$. 
Figure \ref{fig:takasa}(b) shows the phase difference as a function of the mode number 
with $z_\mathrm{m0} = 30$ nm. 
It is shown that the phase difference decreases with increasing mode number $n$.  

\begin{figure}[t]
\centering
\includegraphics[width=80mm]{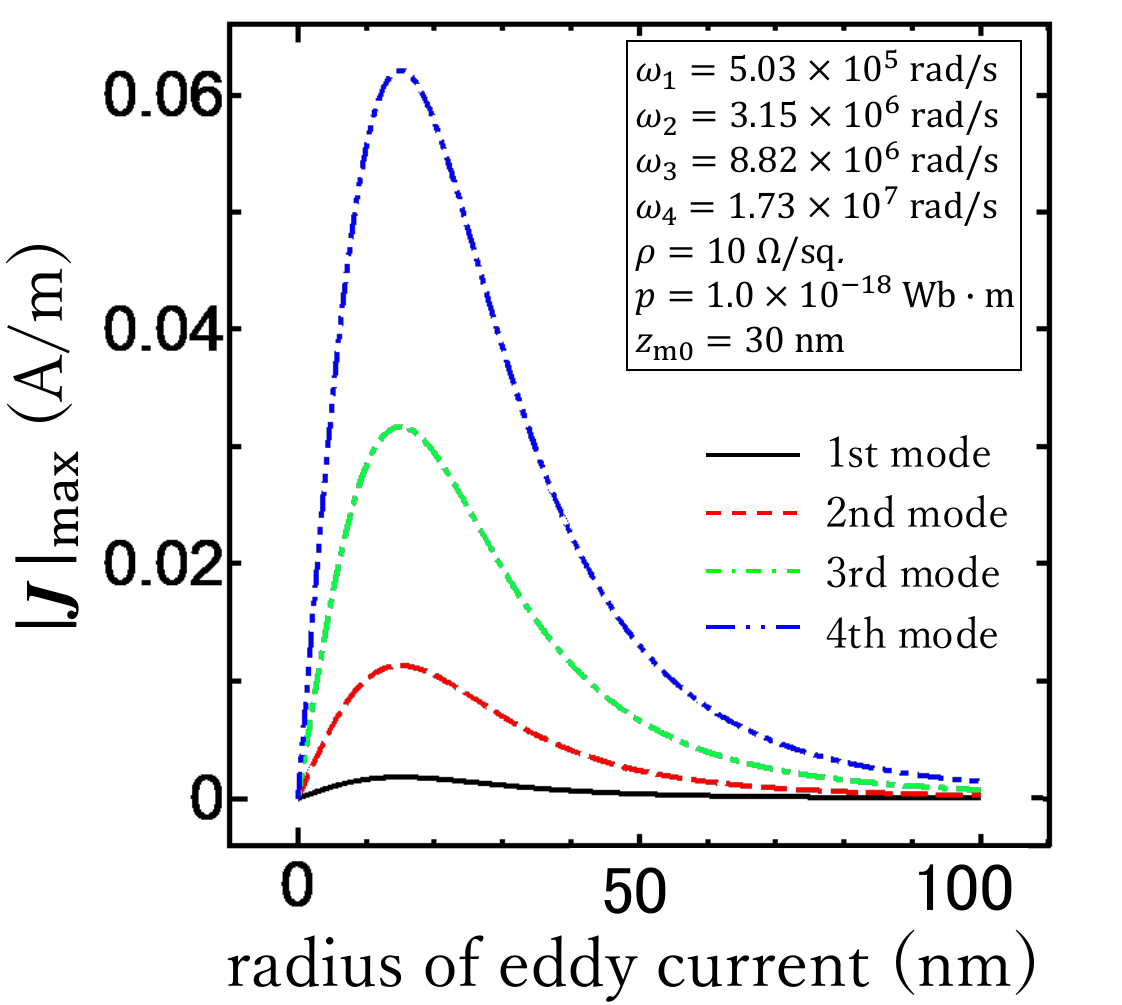}
\caption{\label{fig:denryu}
Eddy-current density of each mode with  $A_n^{(\mathrm{tip})} = 20$ nm  as a function of the radius.}
\end{figure}

\begin{figure}[t]
\centering
\includegraphics[width=75mm]{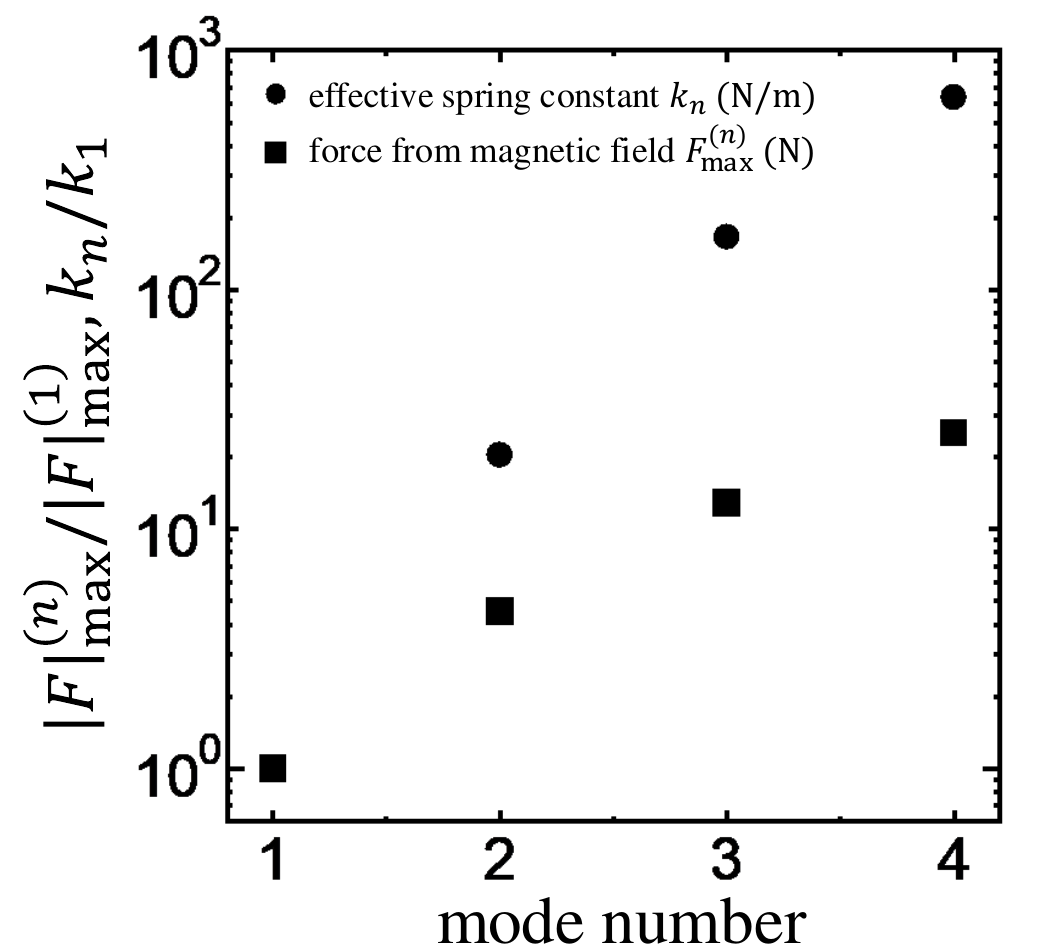}
\caption{\label{fig:raito}Ratio of the maximum force from the eddy current $|F|^{(n)}_{\rm max}$ and 
the effective spring constant $k_{n}$ to those of the lowest mode.
$M_{1}=9.53\times10^{-11}\;{\rm kg},\;M_{2}=4.95\times10^{-11}\;{\rm kg},\;M_{3}=5.15\times10^{-11}\;{\rm kg},\;M_{4}=5.14\times10^{-11}\;{\rm kg},\;\omega_{1}=5.03\times10^{5}\;{\rm rad/s},\;\omega_{2}=3.15\times10^{6}\;{\rm rad/s},\;\omega_{3}=8.82\times10^{6}\;{\rm rad/s},\;\omega_{4}=1.73\times10^{7}\;{\rm rad/s},\;\rho=10\;{\rm \Omega/sq.},\;p=10^{-18}\;{\rm Wb\cdot m},\;A^{({\rm tip})}_{n}=20\;{\rm nm},\;z_{\rm m0}=30\;{\rm nm}$.}
\end{figure}

The reason for the decrease in the phase difference with increasing $n$ can be explained as follows.  
The phase difference of a cantilever feeling the force $F$ as a function of $z$ is generally 
expressed as\cite{ct}
\begin{equation}
\Delta \phi = \frac{Q}{k}\dv{F}{z}, 
\label{eq:30}
\end{equation}
where $k$ is the spring constant. 
This should be valuable in discussing the MFM cantilever oscillating in the $n$th mode.   
The eddy current density $J$ in the $n$th mode is expressed as\cite{n} 
\begin{align}
J & \simeq\frac{3}{4\pi}\frac{p}{\rho}\frac{rz_{\rm m0}}{\{z^2_{\rm m0}+r^2\}^{\frac{5}{2}}}\dv{{u}(L,t)}{t} 
\nonumber\\
& 
= -\frac{3}{4\pi}\frac{p}{\rho}\frac{rz_{\rm m0}}{\{z^2_{\rm m0}+r^2\}^{\frac{5}{2}}}A_n \Psi_n(L) \omega_n  \sin(\omega_n t + \phi_n). 
\end{align}
The force from the eddy current $F_\mathrm{edd}$ in the $n$th mode is expressed as\cite{n}
\begin{align}
F_\mathrm{edd} 
& 
= 
p \dv{}{z}\int_0^\infty \dd{r} \qty[\frac{1}{2}\frac{J}{(z^2+r^2)^{3/2}}]
\nonumber
\\
&
=
-\frac{3}{64\pi}\frac{p^2}{\rho}\frac{1}{z^4_{\rm m0}}A_{n}\Psi(L)\omega_{n}\sin(\omega_n t + \phi_n).\label{eq:fff}
\end{align}
The amplitudes of the sinusoidally oscillating eddy current and force from eddy current are therefore 
\begin{align}
\abs{J}_\mathrm{max} &\simeq   \frac{3}{4\pi}\frac{p}{\rho}\frac{rz_{\rm m0}}{\{z^2_{\rm m0}+r^2\}^{\frac{5}{2}}}A_n \Psi_n(L) \omega_n,\\
|F|^{(n)}_{\rm max} & = \frac{3}{64\pi}\frac{p^2}{\rho}\frac{1}{z^4_{\rm m0}}A_{n}\Psi_n(L)\omega_{n}, 
\label{eq:35}
\end{align}
where $A_n\Psi_n(L)$ is the tip amplitude in the $n$th mode denoted as $A_n^{(\mathrm{tip})}$.    
Figure \ref{fig:denryu} shows $\abs{J}_\mathrm{max}$ with $A_n^{(\mathrm{tip})} = 20$ nm.  
The amount of eddy current increases with increasing $n$ due to the high frequency, 
indicating that the force from the eddy current should increase with increasing $n$. 
This suggests that 
using high-oder modes leads to high sensitivity to measure the resistivity of the non-magnetic material.
In the $n$th mode oscillation case, 
the effective spring constant $k_n \equiv M_n \omega^2_n$ \cite{v} 
increases with increasing $n$. 
Figure \ref{fig:raito} shows the $n$-dependence of the force and the effective spring constant.  
As shown in Fig.\ \ref{fig:raito}, 
the effective spring constant $k_n$ increases faster than the force $\abs{F}_\mathrm{max}^{(n)}$. 
Equation (\ref{eq:30}) with these facts means that the higher-mode leads to poor sensitivity. 

\section{Conclusions}
To investigate the effect of high-order resonance modes of MFM-cantilever oscillation 
on the resistivity measurement sensitivity, 
the equations governing MFM cantilever oscillation in the high-order resonance mode are derived
 considering forced oscillation, dissipation and the force from the eddy current in the material.
The theoretical expression for the phase difference due to the eddy current, which should be 
observed experimentally, is obtained by solving the equation analytically.  
The phase difference decreases with higher modes against expectations.  
This is due to that the effective spring constant increases faster than the force from the 
eddy current.  

\section*{Acknowledgments}
This work was supported by JSPS KAKENHI Grant Number 22H01498.

\section*{Reference}
\bibliographystyle{archive.bbl} 
\bibliography{/Users/okamotokazuma/Desktop/JAPフォーマット/参考文献}

\end{document}